\documentclass{elsart}

\usepackage{graphicx}

\makeatletter
\newcommand{\vect}[1]{\mbox{\boldmath$#1$}}
\newcommand{\txt}[1]{\mbox{\scriptsize#1}}
\newcommand{\bare}{\makebox[0pt]{\phantom{\scriptsize$\overline k$}}}
\newcommand\varcite[1]{{\renewcommand\@cite[1]{##1}\cite{#1}}}
\newcommand{\journref}[5]{#1, #2 #3 (#5) #4.}
\makeatother

\date{}

\begin{document}

\begin{frontmatter}

\title{Interpretation of the Wigner Energy as due to RPA Correlations}

\author{Kai Neerg{\aa}rd}

\address{N{\ae}stved Gymnasium og HF, Nyg{\aa}rdsvej 43,
	 DK-4700 N{\ae}stved, Denmark, {\tt neergard@inet.uni2.dk}}

\begin{abstract}

\noindent In a schematic model with equidistant fourfold degenerate
single-nucleon levels, a conventional isovector pairing force and a
symmetry force, the RPA correlation energy rises almost linearly with
the isospin $T$, thus producing a Wigner term in accordance with the
empirical proportionality of the symmetry energy to $T(T+1)$.

\end{abstract}

\begin{keyword}

symmetry energy, Wigner energy, isovector pairing force, symmetry force,
isobaric invariance, RPA, Goldstone mode

\medskip

PACS-numbers: 21.10.Dr, 21.60.Jz, 27.40.+z

\end{keyword}

\end{frontmatter}

\noindent Nearly symmetric nuclei have an extra binding, the so-called
Wigner energy, that is not described by the quadratic symmetry term in
the semi-empirical mass formula~\cite{MySw66}. This is explained in
various ways in the literature. Counting the even bonds among
supermultiplet degenerate nucleon orbitals, Wigner estimates that the
isospin-dependent part of the interaction energy in the ground state of
a doubly even nucleus is proportional to $T(T+4)$, where $T$ is the
isospin~\cite{Wi}. Talmi proves that for seniority-conserving forces
such as the pairing force acting in a single $j$-shell, this part of the
interaction energy is proportional to $T(T+1)$~\cite{Ta}, and Bohr and
Mottelson point out that this isospin-dependence arises in general from
an interaction proportional to the scalar product of the isospins of the
interacting nucleons~\cite{BoMo}. A symmetry energy proportional to
$T(T+1)$ also fits the empirical masses well~\cite{DuZu}. Myers and
Swiatecki attribute the extra binding of nearly symmetric nuclei to the
interaction of neutrons and protons in overlapping
orbitals~\cite{MySw97}. Shell model calculations with realistic forces
are succesful in reproducing the measured binding energies~\cite{Ca},
whereas with Skyrme forces, no Wigner term appears in
Hartre-Fock-Bogolyubov calculations and only a small one in Hartree-Fock
calculations~\cite{SaWy01}. However, by invoking a particular isoscalar
pairing force that breaks geometric symmetries, Satula and Wyss obtain a
significant Wigner energy in approximately number-projected Bogolyubov
calculations~\cite{SaWy97}. In finite-temperature Bogulyubov
calculations with a Yamaguchi force, R\"opke~{\em et~al.} get at low
temperature in the local density approximation a contribution from
neutron-proton pairing to the binding energies in the $A=40$ isobaric
chain with a maximum at $N-Z=-1$ and a quite irregular dependence on the
asymmetry~\cite{Roe}.

The RPA is the leading order correction to the Hartree-Fock-Bogolyubov
approximation. I have therefore calculated the RPA correlation energy
from the schematic Hamiltonian
\[
		H = H_0- G\vect P^\dagger\cdot\vect P + \frac\kappa2{\vect T}^2,\quad
		H_0 = \sum_{k\sigma\tau}\epsilon_k
		a^\dagger_{k\sigma\tau}a^{\phantom\dagger}_{k\sigma\tau}.
\]
In this expression, the index $k\sigma\tau$ labels orthonormal nucleon
orbitals, and $a_{k\sigma\tau}$ are the corresponding annihilation
operators. $k\sigma$ takes the values $k$ and $\overline k$ so that the
orbital $\overline k$ is obtained from the orbital $k$ by time-reversal,
and $\tau$ is `n' for a neutron orbital or `p' for a proton orbital.
{\vect P} is the pair annihilation isovector, and {\vect T} denotes the
total isospin. The former has the coordinates
\[
		P_x = (-P_{\txt n}+P_{\txt p})/\sqrt2,\quad P_y
  = -i(P_{\txt n}+P_{\txt p})/\sqrt2,\quad P_z = P_{\txt{np}},
\]\[
  P_\tau = \sum_k a_{\overline k\tau}a_{\bare k\tau},\quad
  P_{\txt{np}} = \sum_k (a_{\overline k\txt p}a_{\bare k\txt n}
   + a_{\overline k\txt n}a_{\bare k\txt p})/\sqrt2.
\]
The single-nucleon energy $\epsilon_k$ takes $\Omega$ equidistant values
separated by $\eta$, and $G$ and $\kappa$ are coupling constants.

To describe states with a given number
$
  A_{\txt v} = \sum_{k\sigma\tau}
    a^\dagger_{k\sigma\tau}a^{\phantom\dagger}_{k\sigma\tau}
$
of valence nucleons and a given isospin, I employ the Routhian
\[
  R = H - \lambda A_{\txt v} - \mu T_z.
\]
It is just for convenience that $T_z$ is chosen here as the
isospin-coordinate to be constrained. Since $H$ is isobarically
invariant, one could equivalently constrain the projection of $\vect T$
on any axis in isospace. Following Marshalek~\cite{Ma} I base the RPA on
the Hartree-Bogolyubov (not Fock) self-consistent state derived from
$R$. This is the Bogolyubov vacuum that minimizes $ E_0 - \lambda\langle
A_{\txt v}\rangle - \mu\langle T_z\rangle, $ where
\[
  E_0 = \langle H_0\rangle - \frac{|\vect\Delta|^2}G
    + \frac\kappa2\langle\vect T\rangle^2,\quad
  \vect\Delta = - G\langle\vect P\rangle.
\]
At the minimum one has $\langle T_x\rangle=\langle T_y\rangle=0$. For
large values of $\mu$ a product of neutron and proton BCS states is
expected. Since this state is invariant under the transformation
$\exp(-i\pi(A_{\txt v}/2+T_z))$, I enforce this symmetry, which entails
$\Delta_z=0$. I furthermore assume that both gaps $\Delta_{\txt
n}=(-\Delta_x+i\Delta_y)/\sqrt2$ and $\Delta_{\txt
p}=(\Delta_x+i\Delta_y)/\sqrt2$ are positive, as may always be achieved
by a transformation of the form $\exp(-i(\xi A_{\txt v}+\chi T_z))$.
When $\lambda$ is placed midway between the lowest and the highest
$\epsilon_k$, one then gets $\langle A_{\txt v}\rangle=2\Omega$ and
$\Delta_{\txt n}=\Delta_{\txt p}\equiv\Delta$ for any value of $\mu$ due
to the equidistant single-nucleon spectrum. To speed up the calculation,
I keep $\Delta$ rather than $G$ fixed with the variation of $\mu$. Then
$G$ varies in the case considered by less than .6~\%.

With quasinucleon annihilation operators
$
		\alpha^{\phantom\dagger}_i = \sum_{k\sigma\tau}
		(u^{\phantom\dagger}_{i,k\sigma\tau}a^{\phantom\dagger}_{k\sigma\tau}
		+ v^{\phantom\dagger}_{i,k\sigma\tau}a^\dagger_{k\sigma\tau})
$
defined by
\[
		[\alpha^{\phantom\dagger}_i,R_{\txt{m}}]
    = E^{\phantom\dagger}_i\alpha^{\phantom\dagger}_i,\quad
		E^{\phantom\dagger}_i>0,\quad
  \{\alpha^{\phantom\dagger}_i,\alpha^\dagger_j\}
    = \delta^{\phantom\dagger}_{ij},
\]\[
  R_{\txt{m}} = H_0
    + \Delta(P_{\txt n}^{\phantom\dagger} + P_{\txt n}^\dagger
      + P_{\txt p}^{\phantom\dagger} + P_{\txt p}^\dagger)
    + \kappa\langle T_z\rangle T_z - \lambda A_{\txt v} - \mu T_z,
\]
the RPA Routhian $\tilde R$ is obtained by truncating to second order
the boson expansion of $R$ that results from making in the expressions
for $H_0$, $\vect P$, $\vect T$ and $A_{\txt v}$ the substitutions
\[
  \alpha^{\phantom\dagger}_j\alpha^{\phantom\dagger}_i
    = b^{\phantom\dagger}_{ij} + \cdots,\quad
  \alpha^\dagger_i\alpha^{\phantom\dagger}_j
    = \sum_l b^\dagger_{il}b^{\phantom\dagger}_{jl},
\]
where the boson annihilation operators
$b^{\phantom\dagger}_{ij}=-b^{\phantom\dagger}_{ji}$ satisfy
$[b^{\phantom\dagger}_{ij},b^{\phantom\dagger}_{lm}]=0$ and
$[b^{\phantom\dagger}_{ij},b^\dagger_{lm}]
  = \delta_{il}\delta_{jm}-\delta_{im}\delta_{jl}$. The normal mode
annihilation operators
$B^{\phantom\dagger}_\nu = \sum_{i<j}
  (\phi^{\phantom\dagger}_{\nu,ij}b^{\phantom\dagger}_{ij}
  + \psi^{\phantom\dagger}_{\nu,ij}b^\dagger_{ij})
$
are then given by
\[
  [B^{\phantom\dagger}_\nu,\tilde R]
    = \omega^{\phantom\dagger}_\nu B^{\phantom\dagger}_\nu,\quad
  \omega^{\phantom\dagger}_\nu>0,\quad
  [B^{\phantom\dagger}_\nu,B^\dagger_\rho]
    = \delta^{\phantom\dagger}_{\nu\rho},
\]
and the ground state energy is ~\cite{Ma} $E=E_0+E_2$ with
\begin{eqnarray*}
  E_2\,&=\,&\sum_{k\tau\tau'}
    \left(- G|\langle\alpha_{\overline k\tau'}\alpha_{\bare k\tau}
      \vect P\rangle|^2
    + \frac\kappa2|\langle\alpha_{\overline k\tau'}\alpha_{\bare k\tau}
      \vect T\rangle|^2\right)\\&&
    + {\textstyle\frac12}\left(\sum_\nu\omega_\nu
      - \sum_{i<j}[b^{\phantom\dagger}_{ij},[\tilde R,b^\dagger_{ij}]]
      \right).
\end{eqnarray*}
It may be noticed that when the terms with four quasinucleon
annihilation operators or four quasinucleon creation operators are
removed from the expression for the Hamiltonian $H$ so that the
Bogulyubov vacuum becomes an eigenstate of $H$ (when the operators
$A_{\txt v}$ and $T_z$ are replaced by their eigenvalues in the relation
$H=R+\lambda A_{\txt v}+\mu T_z$), then only the first term in the
expression for $E_2$ survives, and $E_0+E_2$ becomes an exact expression
for the ground state energy. Therefore this expression derived from the
boson expansion is expected to provide a reliable approximation even
though most of the normal modes are not much different from
two-quasinucleon excitations.

The implications of the symmetries of this model for the normal modes
are discussed by Ginoccio and Wesener~\cite{GiWe}. Two Goldstone modes
result from the commutation relations $[A_{\txt v},R]=[T_z,R]=0$.
Furthermore, since $[T_+,R]=\mu T_+$ one normal mode has the frequency
$\mu$. Its annihilation operator is the linear boson part of
$T_+/\sqrt{2\langle T_z\rangle}$, and it becomes a Goldstone mode in the
limit $\mu\rightarrow0$. The degree of freedom of this mode is the
direction of the isospin. In particular the isospin quantum numbers
$M_T=T=\langle T_z\rangle$ may be assigned to the ground state of the
Routhian $R$. The Goldstone modes contribute with the frequency zero to
the expression for the second order energy $E_2$.

\begin{figure}
{\center\includegraphics{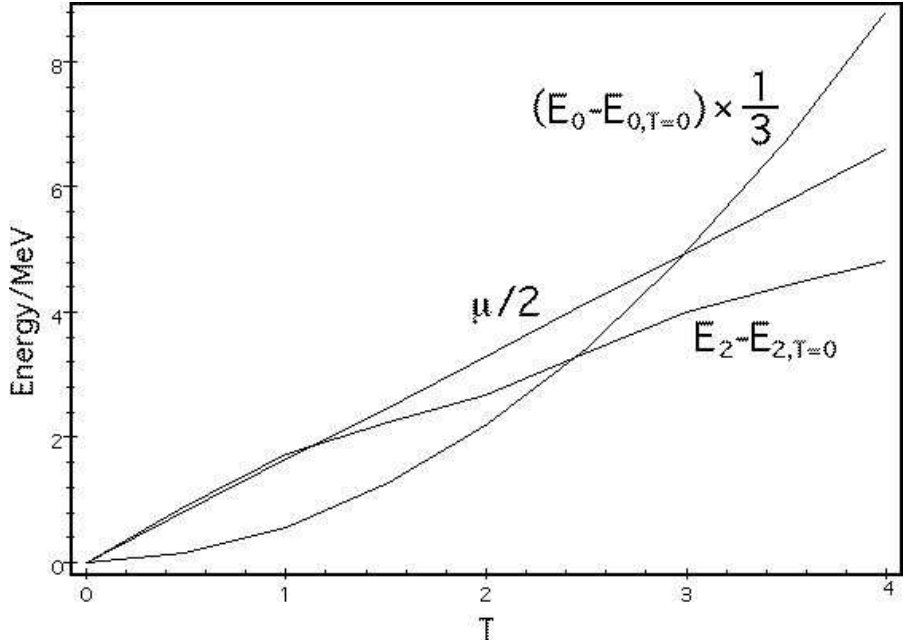}\par}
\caption[]{\label{figure}$E_0-E_{0,T=0}$, $E_2-E_{2,T=0}$ and $\mu/2$ as
functions of $T$. $E_0-E_{0,T=0}$ is scaled by the factor $1/3$.}
\end{figure}

The parameters of the calculation are chosen so as to simulate the
$A=48$ isobaric chain: $\Omega=24$, $\eta=2.1$~MeV, $\Delta=1.7$~MeV,
$\kappa=1.2$~MeV. The result is shown in Figure \ref{figure}.
$E_0-E_{0,T=0}$ depends essentially quadratically on $T$. It is in fact
given in a very good approximation by the expression
$E_0-E_{0,T=0}=\frac12(\eta+\kappa)T^2$ obtained for $\Delta=0$. The
almost exactly quadratic $T$-dependence of $E_0-E_{0,T=0}$ is seen also
indirectly from the linearity of $\mu=dE_0/dT+2(\Delta/G)^2dG/dT$, where
the second term is negligible. $E_2-E_{2,T=0}$ shows a different
behaviour. It rises for $T\approx0$ linearly with $T$ and is in fact in
this limit equal to $\mu/2$. The linearity thus stems from the single
term in the expression for $E_2$ which represents the zero-point energy
of the normal mode with the frequency $\mu$, or, in other words, from
the quantal fluctuation of the isospin.

The rest of the second order energy $E_2$ is faily independent of $T$.
This suggests that the other normal modes are to a large extent
independent of the iso-rotational degree of freedom, or, stated
otherwise, that the iso-rotation is highly collective. The deformation
underlying this collectivity is in the pair field~\cite{FrSh}. Thus with
$\Delta_z=0$ the isovector $\vect\Delta$ is perpendicular to the
iso-rotational axis. So it breaks the iso-rotational invariance with
repect to this axis, and a collective iso-rotation can arise. On the
other hand the contribution to $E_2$ from the non-collective modes
varies from $T=0$ to $T=4$ by almost 2 MeV, so the RPA correlation
energy should be taken into account in a detailed comparison of the
results of Hartree-Fock-Bogolyubov calculations with the empirical
masses.

With $E_0-E_{0,T=0}=\frac12(\eta+\kappa)T^2$ and
$E_2-E_{2,T=0}=\frac12\mu=\frac12dE_0/dT=\frac12(\eta+\kappa)T$ we have
altogether $E-E_{T=0}=\frac12(\eta+\kappa)T(T+1)$, that is, we get the
$T$-dependence of the symmetry energy found in the data. The
Hartree-Fock-Bogolyubov energy expectation value includes only the first
sum in the expression for $E_2$~\cite{Ma}. This sum is cancelled to a
large extent by parts of the second term in the second sum. For the
contributions from the symmetry potential energy $\frac\kappa2\vect T^2$
this cancellation is in fact exact. For $\Delta>0$ the first sum in
the expression for $E_2$ is as a function of $T$ analytic and even at
$T=0$. For $\Delta=0$ it equals, however, $\frac\kappa2T$. Thus it
produces in the absense of pairing a Wigner term, albeit with only
$\kappa/(\eta+\kappa)\approx35$ \% of the full value. Although the
forces there are different, this may explain the experience with
Hartree-Fock-Bogolyubov and Hartree-Fock calculations mentioned in the
introduction.

A Wigner term corresponding to the term $\frac\kappa2T$ in the present
model is actually the only one that may be derived from arguments like
those in References~\varcite{Wi}--\varcite{BoMo}, which are based on the
form of the residual two-nucleon interaction. It is remarkable that with
a deformation one gets also a term $\frac\eta2T$ corresponding to the
`kinetic' part $\frac\eta2T^2$ of the symmetry energy.

The symmetry potential energy $\frac\kappa2\vect T^2$ of the present
model differs from an interaction potential energy
$\kappa\sum_{i<j}\vect t_i\cdot\vect t_j$, where the index $i$ or $j$
labels the individual nucleons, by the term $\frac38\kappa A_{\txt v}$,
which depends only on the number $A_{\txt v}$ of valence nucleons. This
interaction favours isoscalar nucleon pairs. It is included in the
Hamiltonian in order to get a realistic symmetry energy coefficient. It
is in fact well known and discussed in detail by for example Bohr and
Mottelson~\cite{BoMo} that the kinetic term accounts for only a part of
the empirical symmetry energy coefficient. Bohr and Mottelson consider a
contribution to the single-nucleon potential energy corresponding to a
two-nucleon potential of the form $\kappa\vect t_1\cdot\vect t_2$. Thus
the form of the symmetry potential energy of the present model is
adopted from their discussion. This symmetry force is inessential,
however, for the formation of the Wigner energy, which results, as it
was seen, from the collectivity of the iso-rotation due to the isobaric
non-invariance of the isovector pair field.

I am indebted to Stefan Frauendorf for drawing my
attention to the issue of the Wigner energy and for discussions of the
matter.

\end{document}